\documentclass[fleqn,usenatbib]{mnras}
\usepackage{newtxtext,newtxmath}
\usepackage[T1]{fontenc}
\usepackage{ae,aecompl}
\usepackage{tabularx}

\usepackage{graphicx}	
\usepackage{amsmath}	
\usepackage{amssymb}	
\usepackage{csvsimple}

\title[UV upturn in Abell~1689]{Evolution of the UV upturn in cluster galaxies: Abell 1689}

\author[S.~S. Ali et al.]{S.~S. Ali$^1$\thanks{Email: s.ali@bristol.ac.uk}, M.~N. Bremer$^1$, S. Phillipps$^1$, R. De Propris$^2$ \\
$^1$H.H.~Wills Physics Laboratory, University of Bristol, Tyndall Avenue, Bristol, BS8 1TL, UK. \\
$^2$FINCA, Turku Observatory, Turku, Finland. \\}

\date{Accepted XXX. Received YYY; in original form ZZZ}

\pubyear{2018}

\begin{document}
\label{firstpage}
\pagerange{\pageref{firstpage}--\pageref{lastpage}}
\maketitle

\begin{abstract}
We have measured the strength of the UV upturn for red sequence galaxies in the Abell~1689 cluster at $z=0.18$, reaching to or below the $L^*$ level and therefore probing the  general evolution of the upturn phenomenon. We find that the range of UV upturn strengths in the population as a whole has not declined over the past 2.2 Gyrs. This is consistent with a model where hot horizontal branch stars, produced by a Helium-enriched population, provide the required UV flux. Based on local counterparts, this interpretation of the  result implies Helium abundances of  at least 1.5 times  the primordial  value for this HB population, along with high formation and assembly redshifts for the galaxies and at least a subset of their stellar populations.
\end{abstract}



\section{Introduction}

The spectral energy distributions 
of early-type galaxies show a sharp rise in the ultra-violet at $\lambda$ $<$ 2500 \AA, a phenomenon known as the UV upturn (hereafter upturn), first discovered through UV observations using OAO (\citealt{code1979}) and IUE (\citealt{bertola1982}). Given that the stellar populations of these galaxies are mostly old and metal-rich, with little or no evidence of recent star formation, the likely sources of the upturn are a population of hot horizontal branch (HB) stars \citep[e.g.][]{greggio1990,bressan1994,dorman1995}. This explanation is supported by  spectroscopy and imaging of local galaxies \citep[e.g.][]{brown1997,brown1998b,brown2000b}. The stars in this population can only evolve onto the hot HB (generally beyond the RR Lyrae gap) in these old and metal-rich galaxies providing that they have an unusually  high helium abundance ($Y$). Low-metallicity stars also end up on the blue HB in local globular clusters and this along with the age of the populations have also been considered as possible explanations for the upturn (\citealt{yi1998}; \citealt{lee1994}; \citealt{park1997}). However, the presence of a significant fraction of low metallicity stars in these massive galaxies is already excluded by the observation of their strong metal lines which imply that the dominant population is solar or super-solar in abundance. Furthermore, our observations in \cite{ali2018}, hereafter Paper I, show that the contribution of the upturn population spectral energy distribution, if parameterised as a black body,  is more prominent and hotter in the more massive and more metal rich galaxies. This is contrary to what one would expect in a low-metallicity scenario. A low metallicity model would imply that a blue HB population should only appear at the lowest redshifts, and certainly be missing in observations of even moderate-redshift galaxies, contrary to previous work which detected upturns in the most massive galaxies over the past $\sim 5$~Gyrs, e.g. in the work of \cite{brown1998a,brown2000a,brown2003}. The work we descibe below confirms that the upturn remains to at least $z\sim 0.2$ in the general cluster early type population. Similarly, older stellar populations may also produce hot HBs without Helium enrichment, but we discuss below how this is unfeasible given the age of these galaxies and the constraints of cosmological parameters.

UV luminous HB populations  are  observed in a subset of Galactic globular clusters  \citep[e.g.][]{piotto2005,piotto2007}, which are often interpreted as having high helium abundances. While there could be other origins for such a UV bright population, such as close binaries \citep{han2007} and stars with excessive mass loss (as a function of metallicity) on the red giant branch \citep{bressan1994,yi1997,yi1998,yi1999}, these  are not expected in globular clusters and so are unlikely to account for their UV luminous sub-populations. Furthermore, mass loss on the red giant branch is not observed to depend on metal abundance in local clusters \citep{miglio2012,salaris2016}, and the binary fraction appears to decrease with metallicity in the bulge \citep{badenes2017}, the reverse of what would be required to produce an upturn in these otherwise metal rich galaxies. The problem of accounting for a hot HB population without an enhanced helium abundance is exacerbated in early-type galaxies relative to those in globular clusters because the higher metallicity of stellar populations in the galaxies would otherwise act to cool any HB population relative to those in globular clusters. See Paper I for a more extended discussion of this.

Assuming that the upturn is produced by a helium enhanced HB population, it should evolve strongly with age through its sensitivity to the HB morphology and the time-dependence of the main sequence turnoff mass. Given that the more massive stars never reach the blue HB, there is a  limit to the time since a stellar population formed before which a hot HB could appear, irrespective of the level of Helium enhancement. In some of the most metal-rich clusters in our Galaxy, only the $\sim 10\%$ of stars with high $Y$ values populate the hot HB, although as much as 2/3 of the stellar population may have non-cosmological $Y$ abundance \citep{tailo2017}. Because the upturn switches on `rapidly' as the turnoff mass reaches a value that allows HB stars to form, it is potentially a sensitive probe of the epoch of galaxy formation, accessible by observations even at moderate redshifts \citep[ e.g. out to $z\sim0.6$,][]{tantalo1996,chung2017}.

\begin{table}
\label{table}
\begin{filecontents*}{table.csv}
Filter,Proposal ID,PI,Total exp. time/s
ACS F475W,9289,Ford,9500
ACS F625W,9289,Ford,9500
WFC3 F225W,12931,Siana,27710
WFC3 F275W,12201,Siana,73360
WFC3 F336W,12931,Siana,33075
\end{filecontents*}
\csvautotabular{table.csv}
\caption{Table giving details of the images extracted from HLA for each of the HST filters. Further information regarding the ACS and WFC3 images can be found in \protect\cite{mieske2004} and \protect\cite{alavi2016} respectively.}
\end{table}

Given that HB stars are at a post-MS evolutionary stage of their life cycle, the upturn is a direct probe of the properties of this old stellar population that potentially formed at a very high redshift ($z_f\sim4$ and above as suggested by, e.g. \citealt{brown1998a,brown2000a,brown2003}). Most previous work has concentrated on very bright early type galaxies in clusters because of limitations in telescope aperture or detector sensitivity. \cite{brown1998a, brown2000a, brown2003} studied the evolution of the upturn in  brightest cluster  galaxies at $z\sim 0.3-0.6,$ and found that the rest-frame $FUV-V$ colour in general reddens with redshift in these galaxies. \cite{ree2007} also performed a similar study on brightest cluster ellipticals at $z<0.2$ and found a similar fading of the upturn with redshift. More recently \cite{boissier2018} found that the upturn exists in a sample of several tens of BCGs out to $z\sim 0.35$ behind the Virgo cluster. 

In Paper I we explored the strength of upturns in the red sequence galaxy populations of local clusters including Coma. In this and subsequent papers we explore the evolution of the upturn in cluster red sequence galaxies over a range of redshifts and for a range of luminosities, rather than just the very brightest galaxies.  Here we concentrate on the galaxy population of Abell 1689 at $z=0.18$, which at the initiation of this project was the only cluster at $z\sim 0.2$  with suitable archival HST ACS and WFC3 imaging data in the required bands and of sufficient depth to reach   below the $L^*$ point in the luminosity function. This allows us to directly compare the spread in the upturn strength of galaxies in Abell 1689 to those in  local clusters (Paper I) down to $L^*$ and beyond. Otherwise, we take this cluster as a typical exemplar of massive clusters at this redshift.

All magnitudes quoted are in the AB system, and the cosmology  assumes $h=0.7,\Omega_m=0.3,\Omega_\Lambda=0.7$. Galactic extinction corrections were made using the extinction maps from \cite{schlafly2011}.

\begin{figure}
\includegraphics[width=0.5\textwidth]{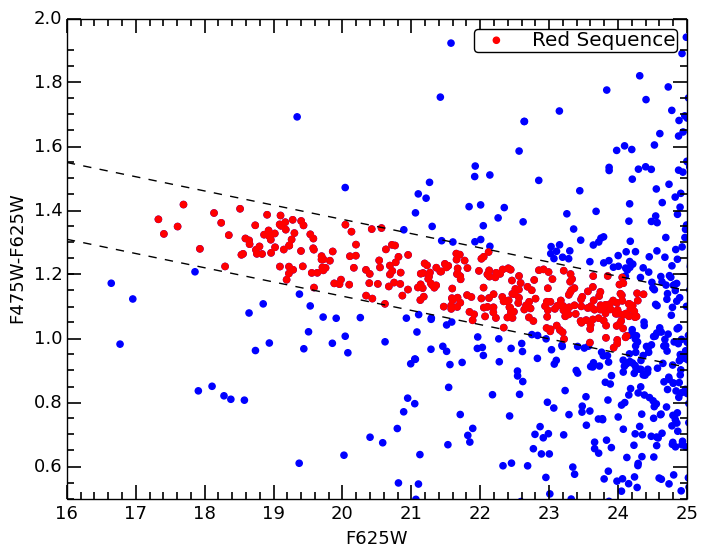}
\caption{Optical F475W-F625W vs F625W colour-magnitude diagram. The red sequence is denoted by red filled circles and have photometric uncertainty of <0.05 magnitudes in their optical colour.}
\label{fig:g-r}
\end{figure}

\section{Data \& Photometry}

For Abell 1689, suitable optical and UV imaging data is available from the HST archive. 
The  optical data consist of images taken with the Hubble Space Telescope (HST)'s Advanced Camera for Surveys (ACS) through the F475W and F625W filters, corresponding approximately to the rest-frame $B$ and $V$ filters at this redshift\footnote{See \cite{avila2017} for details of the ACS.}. We extract archived images in  both filters from the Hubble Legacy Archive (HLA). The program IDs and PI identifications are reported in Table \ref{table}, together with exposure times and other relevant information.

The ACS F475W and F625W are level 3 HLA mosaics, which are combined images from multiple HST visits covering a large contiguous area of the sky. Level 3 data are produced through the DrizzlePac pipeline which removes geometric distortions, positions the image North up, corrects for sky background variations, removes cosmic rays and combines and projects multiple images on to one common frame with the same pixel scale as the detector. Details of the whole drizzling process is given in the DrizzlePac handbook (\citealt{gonzaga2012}). As Level 3 data have already been pre-processed by the pipeline, photometry could be carried out on them without any need for further corrections to the images.

\begin{figure*}
{\includegraphics[width=0.49\textwidth]{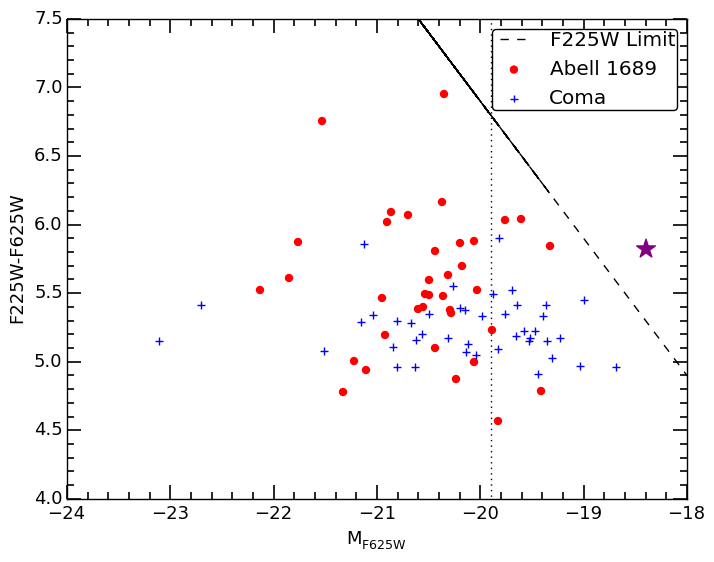}}
{\includegraphics[width=0.49\textwidth]{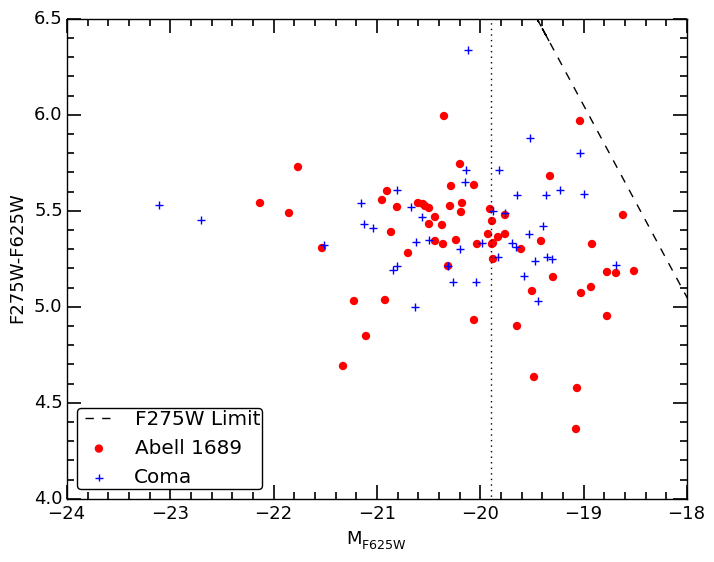}}
\caption{$F225W-F625W$ (left) and $F275W-F625W$ (right) {\it vs} $M_{F625W}$ colour-magnitude diagrams for the Abell 1689 red sequence galaxies (red filled circles). Photometric uncertainties in colours are <0.15 magnitudes. The purple starred data point represents the F225W-F625W colour of the stacked galaxies between $\sim M_{F625W}=$-19 and -18. Also plotted (blue crosses) is photometry of Coma red sequence galaxies from Paper I obtained in similar rest-frame bands ({\it i.e.} UVOT $UVW2-V$, left, and GALEX $NUV-V$, right) in the same  absolute magnitude range as for the Abell 1689 sample. The vertical dotted line denotes the $L^*$ point for Abell 1689. See text for details.}
\label{fig:uv}
\end{figure*}

We used SEXTRACTOR \citep{bertin1996} to carry out aperture photometry (through a metric 7.5kpc diameter aperture) on the F475W and F625W images, deriving also the \cite{kron1980} magnitudes in each band. The red sequence for this cluster is clearly visible in Fig. \ref{fig:g-r}. Because there is no complete redshift survey for this cluster, we make the reasonable assumption here that, all bright resolved galaxies falling in the region of parameter space coincident with the cluster red sequence are cluster members. However, CLASH-VLT spectroscopy of rich clusters at similar redshifts indicates that $>95\%$ of apparent red sequence galaxies at these magnitudes are genuine cluster members \citep{depropris2016}. We do not consider red sequence galaxies fainter than F625W$>24$ as the possibility of contamination by foreground or background objects significantly increases at this point. Given the tight red sequence down to our selection limit, contamination from non-cluster members is therefore unlikely to be significant.

Ultraviolet observations were made using using the Wide Field Camera 3 (WFC3) with the F225W, F275W and F336W filters of the UVIS channel, corresponding to rest frame 1900\AA, 2300\AA\ and 2850\AA\ respectively\footnote{Details of the WFC3 detector is given in \cite{dressel2017}.}.

The WFC3 images through these filters are provided by the archive as  Level 2 data, which are identical to  Level 3 data in how they are processed with the  exception  that the combined images are limited to the same HST visit. In order to achieve the maximum depth possible for each band, we aligned and combined all Level 2  frames in each filter using the \textit{imalign} and \textit{imcombine} functions within \textit{IRAF}. 

The different pixel scales of the optical and UV images precluded the use of SEXTRACTOR in dual mode to obtain UV photometry in the manner as for the optical data.  Instead, we used \textit{IRAF}'s \textit{apphot} package to place apertures of fixed 7.5kpc diameter on to the RA and DECs of all the red sequence objects determined from the optical data onto  the F225W, F275W and F336W images to measure their corresponding magnitudes in these   bands. We then made a 5${\sigma}$ signal to noise cut on our detected objects and checked them by eye to ensure they are indeed real. In total we find a sub-sample of 37 optically-selected red sequence galaxies that are detected in every UV band. Of these, approximately half had redshifts tabulated in NED\footnote{ned.ipac.caltech.edu} and they were all confirmed cluster members. These  represent the 37 optically-brightest galaxies of a sample of 176 red sequence galaxies within the field-of-view of the UV frames. The non UV-detected galaxies had upper limits to their UV-optical colours which do not constrain the presence or otherwise of an upturn. The detected objects typically displayed  visually identifiable rest-frame UV emission only over the  central $\sim1$ arcsec, simply reflecting the relative surface brightness sensitivity to $z=0.18$ ellipticals of the HST data in the different bands. At the redshift of Abell~1689, we detect F225W emission from all ellipticals with optical luminosities down to the $\sim L^*$ level (\citealt{banados2010}).

\section{Results}
\subsection{UV-to-optical colour-magnitude diagrams}
We plot in Fig. \ref{fig:uv} (left) the $F225W-F625W$ (rest-frame $\sim1900-V$) colours against the absolute F625W (rest-frame $V$ band) magnitudes. There is a spread of approximately 2 magnitudes in this colour. While not a conventional means of measuring the strength of the upturn, this colour has similar sensitivity  to the upturn as the standard $1550-V$ (GALEX $FUV-V$) colour more normally used to measure its strength. Consequently,  the spread of $\sim2$ magnitudes in $1900-V$ can be compared to that in the $1550-V$ colour of red sequence galaxies found in Coma, Fornax, Perseus and Virgo clusters (Paper I, \citealt{boselli2005}). We can also make a more direct comparison with the UVOT $UVW2-V$ ($\sim1900-V$) colours of Coma galaxies from Paper I for the same luminosity range as our Abell~1689 sample, where the galaxies show a somewhat smaller scatter of $\sim1.5$ magnitudes. It is important to note that while the {\it spread} in the $UVW2-V$ and $F225W-F625W$ colours between the two clusters may be comparable, we can not as easily compare the actual colour values. This is due to the inherent differences in the shape and bandwidths of the HST WFC3 F225W and UVOT UVW2 filter responses, which have FWHMs of 500\AA\ and 657\AA\ respectively. These differences lead to an underlying variation in the measured colours independent of any upturn.

We also stack the non-detected galaxies in the F225W band between $\sim M_{F625W}=$-19 and -18 to get a detection and determine the average $F225W-F625W$ colour of the galaxies just beyond the detection limit. This is plotted as the purple star in Fig. \ref{fig:uv} (left). As can be seen, the colour is consistent with some of the redder detected galaxies in our sample, which is to be expected given that the upturn tends to get weaker with decreasing mass/luminosity (\citealt{boselli2005}; \citealt{smith2012}).

It should be noted that our sample of Abell~1689 galaxies does not contain the two brightest (central) galaxies. This is due to the final UV image of the cluster having a 'dead' zone at its centre, resulting from  the range of rotations and offsets of individual data frames not fully compensating for the gap between individual detectors. As seen from previous studies, locally the largest galaxies -- particularly BCGs -- tend to have some of the bluest UV-optical colours, and as such the strongest upturns (\citealt{burstein1988}), though the results of \cite{brown1998a,brown2000a,brown2003} and \cite{ree2007} noted earlier may indicate some evolution for these objects.

We also plot in Fig. \ref{fig:uv} (right) the $F275W-F625W$ (rest frame $\sim2300-V$) colours of our galaxies against the absolute F625W magnitudes. While not as sensitive to the upturn as the GALEX $FUV-V$ used in Paper I, previous studies of red sequence galaxies (\citealt[e.g.] {schombert2016}) have shown that the $NUV-V$ colour is also affected quite strongly by the hot HB stars that are the likely source of the upturn. In the discussion section we further justify this choice, showing that the $NUV-V$ colour probes the upturn and does not depend significantly on the underlying stellar populations. Since in the  rest frame of Abell~1689,  the HST $F275W-F625W$ colour is very similar to that of the the GALEX $NUV-V$ measured for the sample of Coma galaxies with the same absolute magnitude range as that for Abell~1689, these results can be directly compared, as seen in Fig. \ref{fig:uv} (right). Both the Abell~1689 and Coma galaxies show a near identical spread of $\sim1.5$ magnitudes and have very similar $NUV-V/F275W-F625W$ colours. This immediately suggests that if the range in this colour is due to upturns of varying strengths, the range in upturn strength has not evolved drastically over the past $\sim2.2$ Gyrs. In addition, it is clear that this is true for the {\it whole} population of galaxies, and not just the brightest galaxies as seen in previous studies \citep{brown1998a,brown2000a,brown2003}.

At first glance, these results seem to be in disagreement with those of \cite{ree2007}, who found that the upturn fades with redshift in the range of $z=0-0.2$. However, this previous study was performed only on the BCGs. When our much larger sample of elliptical galaxies down to $L^*$ (and beyond) is considered, we find that strength of the upturn phenomenon does not appear to evolve significantly out to $z=0.2$.

\subsection{UV-to-optical SEDs of Abell~1689 galaxies}
\begin{figure}
\includegraphics[width=0.5\textwidth]{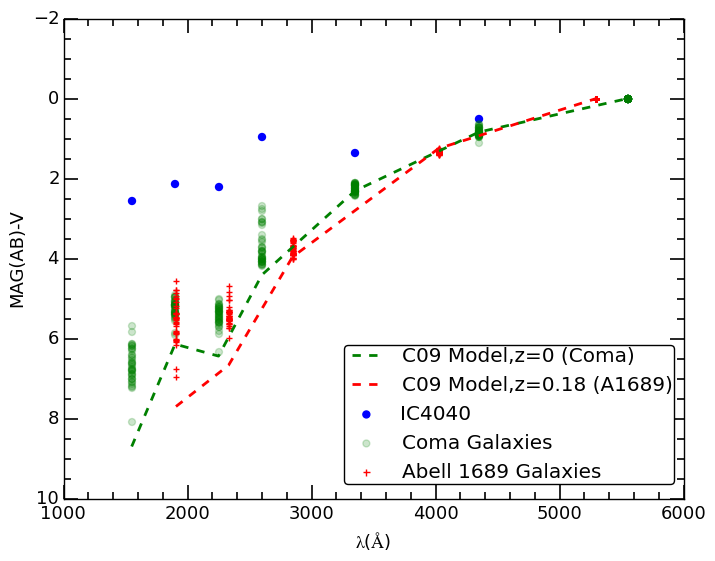}
\caption{UV to optical SEDs of Abell 1689 red sequence galaxies, IC4040 (a starforming galaxy) and an old metal-rich SSP (with no star formation) from the C09 model. Also plotted are similar UV to optical SEDs of Coma red sequence galaxies for comparison. Photometric uncertainties in colours are <0.15 magnitudes.}
\label{fig:abell_sed}
\end{figure}

We show in Fig. \ref{fig:abell_sed}  the photometry of the 37 galaxies in our sample that were detected in all WFC3 and ACS bands. The observed  F225W-F625W, F275W-F625W, F336W-F625W and F475W-F625W colours  are plotted  against the equivalent rest-frame central wavelength of each filter, to create  UV-to-optical SEDs that span the wavelength range between $\sim1500$\AA\ and $\sim5500$\AA. Even though  the SEDs of the  Abell~1689 galaxies are sampled by fewer points in the UV than for those of the lower redshift Coma galaxies from Paper I, we  plot both sets of SEDs in order to compare as many data points between the two clusters as possible. We also plot  the SED of IC4040, a typical star-forming galaxy from the Coma cluster. The SED of this galaxy is much flatter between the UV and optical than those of  red sequence galaxies, which is a result of the fundamental difference between the source of UV luminosity in star-forming and quiescent galaxies. 

Also included in Fig. \ref{fig:abell_sed} is the  the SED of a "red and dead" SSP from \cite{conroy2009} (C09 henceforth) with solar metallicity and a formation redshift of $z_f=4$, which represents a system completely dominated  by a conventional old stellar population and therefore having a very weak UV output. The same C09 model is used for both dashed curves. The offset between them is due to its convolution with the appropriate (rest-frame) filter bandpasses for the two sets of observations. Although other models for synthetic SSPs exist, the advantage of using C09 is that it specifically does not attempt to model in an ad-hoc fashion any post main-sequence stellar population that contributes to a galaxy's UV emission (see Paper I).

The SEDs of IC4040 and C09 represent the two opposite extremes of ongoing star formation and no star formation at all, which can then be compared to our sample of red sequence galaxies.  We can form intermediate SEDs between these two extremes by adding in a proportion of the IC4040 SED to that of the C09 model. Given that the range in optical colours of our selected Abell~1689 red sequence galaxies is strongly limited (covering a range of no more than 0.1 magnitude in F336W-F625W), this constrains the maximum proportion of IC4040 SED that can be included. Adding this maximal amount to the C09 SED leads to a combined SED that  is redder in  UV-optical colours than those measured for our red sequence galaxies, implying that the UV flux is dominated by emission not arising from ongoing star formation, as similarly found in Paper I.

As already noted, the galaxies in our Abell~1689 sample have a very tight spread of $\sim0.1$ magnitudes in their optical colour. However, at wavelengths shorter than 3000\AA, the spread in the colours gradually become larger, increasing  to several magnitudes at the shortest wavelength (1900\AA). Clearly, the large scatter seen in the UV colours is absent in the optical colours. This suggests that the sub-population of stars responsible for the  scatter in the UV has little to no effect on the optical output of these galaxies. Furthermore, the C09 SSP fits our optical data extremely well, but in the UV the same model has colours at least as red as the reddest of our observed galaxies. Given that the C09 SSP replicates the contribution of the conventional old stellar population, this indicates that such a population has a very minor effect on the overall UV emission of red sequence galaxies, as one would expect. Ultimately, the comparison of our results with models suggest (as in Paper I) that these red sequence galaxies are formed of a majority old stellar population that emits strongly in the optical, superimposed with  a similarly old sub-population of (likely) He-enhanced hot HB stars that emit strongly in the UV. The large range  in the UV-optical colours reflects  the different strengths of the upturn in different galaxies, which in turn appears to be  linked directly to the temperature and number of hot HB stars within the galaxies (see Paper I).

\begin{figure}
\includegraphics[width=0.5\textwidth]{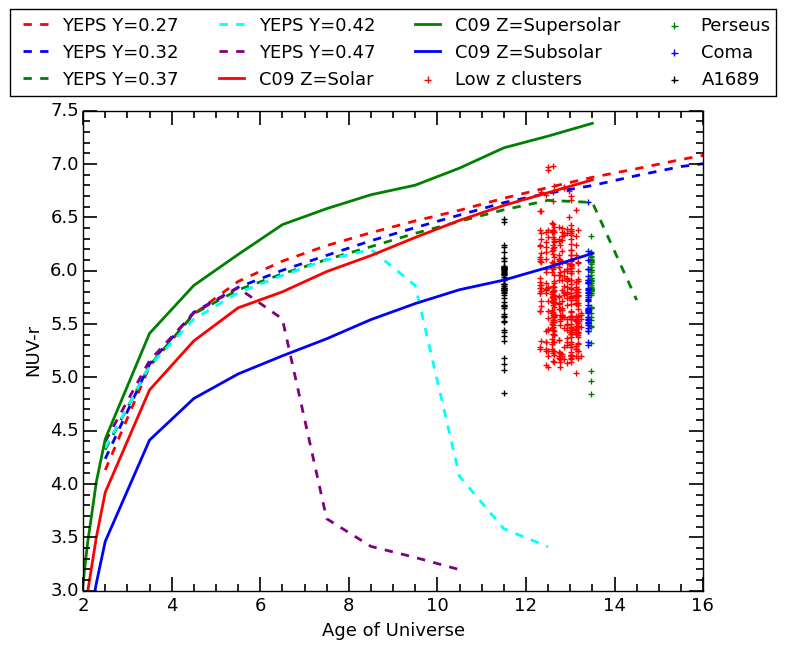}
\caption{YEPS spectrophotometric models (assuming $z_f=4$ and $Z=Z_{\odot}$) showing the evolution of the GALEX NUV-r colour over the age of the universe for a range of Helium abundances. Also included in the plot are the predictions from the C09 model with the following metallicities: $Z$=\(Z_\odot\), 0.56\(Z_\odot\) and 1.78\(Z_\odot\). Plotted on top are the NUV-r colours of Coma, Perseus and Abell 1689 red sequence galaxies. Photometric uncertainties in colours are <0.15 magnitudes.}
\label{fig:yeps}
\end{figure}

When compared with Coma, the spread in the optical colours of Abell~1689 red sequence galaxies is similarly tight, $\sim0.1$ magnitudes. As discussed previously, the spread in the $F275W-F625W$ (Abell~1689) and GALEX $NUV-V$ (Coma) colours are also remarkably similar. In the case of the $F225W-F625W$ (Abell~1689) and UVOT $UVW2-V$ colours (Coma), the exact values between the two clusters can not be directly compared due to the inherent difference in the bandpasses of the HST F225W and UVOT UVW2 filters, which is illustrated by the large difference in the $\sim1900-V$ colour of $\sim1.5$ magnitudes in the underlying C09 model when convolved through each filter. Such a large difference in the model colours does not exist (nor was it expected to exist given the bandpasses) for any of the other data points. 

As discussed in Paper I, the Coma galaxies have an unexpectedly large spread of $2$ magnitudes in the $UVW1-V$ ($2600-V$) colour, larger than all other shorter wavelength UV colours. It was surmised that this may be an issue with a subset of the UVOT UVW1 dataset, rather than the true photometry of the galaxies. The majority of the Coma galaxies could be well fitted  by a C09 SSP as shown in Fig. \ref{fig:abell_sed} combined with a blackbody of a specific temperature and normalisation to account for the UV emission. Those galaxies that could not be fit with such a 2-component model were the ones with the unusually blue $UVW1-V$ colours ($\leqslant3.8$ mags) - the discrepant $UVW1$ magnitudes specifically degraded any fit. From our current data we note that the F336W-F625W (rest frame $\sim2800-V$) colour of the Abell 1689 galaxies should be  closely related to that of the  $UVW1-V$ ($2600-V$) colour for the Coma galaxies. The former colour has a smaller range than the latter, and when those objects in Coma that are not well-fitted by a two component model are excluded, the ranges become comparable. This lends credence to the argument in Paper I that the overly large spread in $UVW1-V$ in the Coma galaxies is caused by an issue with a subset of the UVW1 data.

\section{Discussion}

As discussed in the introduction and in Paper I, we can interpret our results in terms of the presence of a Helium enriched HB population within the galaxies giving rise to their upturns. Recently \cite{chung2017} published synthetic photometry for a series of Simple Stellar Populations with varying levels of Helium abundance ($Y$) and metallicity ($Z$) as a function of time since burst (referred to here as YEPS models, after the name of the project). We can use these models to interpret our results and place constraints on the age and/or Helium enrichment of any population giving rise to the upturn.

Fig.\ref{fig:yeps} shows the evolution of the GALEX $NUV-r$ colour with lookback time as given by the YEPS spectrophotometric models for a range of He-enriched populations as dashed lines. We note here that the photometry from the YEPS models are tabulated by \cite{chung2017} for a range of $Y_{ini}$, which is defined as the initial Helium content of the SSP at $Z=0$. This parameter is related to the Helium abundance through the following equation: $Y=Y_{ini}+2\times \Delta Y/\Delta Z \times Z$. We adopt the values of $\Delta Y/\Delta Z=2$ and $Z=0.02$ (solar metallicity) to calculate the $Y$ for our models as shown in Fig. \ref{fig:yeps}, and use these values in the discussion henceforth. We also assume a formation redshift of $z_f=4$. A later formation redshift shifts these evolutionary tracks to the right, trading decreased age for increased $Y$ at a given colour. Age and $Y$ are degenerate in these colours for these models because higher $Y$ populations have shorter main sequence lifetimes. We then plot  the $NUV-r$ colours of Coma and Perseus from Paper I and low redshift red sequence galaxies from twenty further $z<0.1$ clusters (described in Ali et al., in prep). We also add  the $F275W-F625W$ colours of the Abell~1689 galaxies, suitably k-corrected to the rest-frame $\sim NUV-r$.  The overall distribution of points demonstrates a lack of  evolution for  the UV upturn out to  $z \sim0.2$ (with a $\sim2.2~Gyr$ range in lookback time) - the range in the $NUV-r$ of $1\sim1.5$ mags, which is a measure of the upturn strengths, is consistent between all the clusters in the redshift range being surveyed. 

The observed range in the $NUV-r$ colour is also consistent with the predictions from the YEPS models. The reddest objects can be explained as those with  little or no contribution from a significantly He-enhanced HB population. Conversely, the bluest require an additional population with  $Y>0.37$ if the formation redshift was $z_f\sim 4$ or higher, and  an even stronger enrichment if the  population formed at lower redshift. It should be noted that the model colours shown in Fig. \ref{fig:yeps} are for uniformly He-enhanced populations of a given $Y$ and fixed $Z=Z_{\odot}$. In reality, our observed data are for real red sequence galaxies that have a majority "red and dead" population with a standard Helium fraction that is close to primordial, combined with a sub-population of He-enhanced stars with a range of $Y$. As such, our observed colours do not (and should not) reach the bluest colours shown by the models. Because of the age-$Y$ degeneracy, it is possible in theory to have populations with very high $Y$ values (as seen in the $Y=0.42$ and $Y=0.47$ models), which become UV-bright after $\sim7$ and $\sim4$ Gyrs main sequence lifetimes respectively. These populations could therefore form at much lower redshifts ($z\sim1$). However, given the fact that the He-enhanced sub-population is not an insignificant fraction (of order 10\%) of the galaxies' overall stellar population (Paper I) and that most cluster red sequences appear to be already established by a redshift of $z\sim 1-2$ (e.g. \citealt{newman2014}), any appreciable star formation would seem to have been completed at much higher redshifts (i.e. more than 10 Gyrs ago).

The upturn typically becomes obvious as a separate component of the SED at $\le 2500$\AA\  and  strengthens to at least $\sim1500$\AA. Although measurements in the NUV band (rest frame $F275W$ for Abell~1689) can be significantly influenced by the upturn, we note that it is also partially affected by the output from the main sequence. Consequently it is potentially sensitive to variations in  the age and metallicity of the entire stellar population. To test whether variation in these main sequence parameters can explain the observed range in the $NUV-r$ colour seen in our clusters, we plot in Fig. \ref{fig:yeps} alongside the YEPS models, the evolution of $NUV-r$ against the age of the Universe as given by the C09 models (which have no upturn component) for three different metallicities - $Z$=\(Z_\odot\), 0.56\(Z_\odot\) and 1.78\(Z_\odot\) (i.e. solar, sub-solar and super-solar) with $z_f=4$. As can be seen from the plot, the C09 model with solar metallicity matches closely the YEPS model with $Y=0.27$, since the latter has no upturn compared to the models with higher $Y$. The super-solar and sub-solar models have $NUV-r$ colours that are $\sim0.5$ mags redder and bluer than the solar metallicity model respectively. Since the galaxies we probe are almost all at or above the $L^*$ point, the majority will have solar or super-solar metallicities. This is supported by \cite{price2011} who showed that all Coma galaxies above $L^*$ (i.e. the ones plotted in Fig. \ref{fig:yeps}) have metallicities between $Z$=1-2\(Z_\odot\). In Paper I, we also attempted to fit blackbodies+C09 models with sub-solar metallicities to the SEDs of the aforementioned Coma galaxies, but these all gave poor fits (see Paper I for further details). 
Thus the $NUV-r$ colours given by the $Z$=0.56\(Z_\odot\) metallicity C09 model is unlikely to be appropriate for any of the galaxies in our sample. In any event, the model is still not blue enough to account for the entire range of $NUV-r$ colours observed in these clusters even if the entire stallar population of a galaxy has this metallicity. If such a population is mixed in with a super-solar population (in order to satisfy all previous evidence for higher overall metallicities), the mismatch in the range of colours is even worse. 

As noted earlier, the age of the main sequence population is the third parameter besides metallicity and Helium enhancement to potentially have a significant effect on the $NUV-r$ colour. As can be seen from the time-evolution of the colours of all C09 models, the change in the $NUV-r$ colour is $\sim0.1$ per Gyr. Recent studies by \cite{jorgensen2017} have found a lack of any appreciable star formation in cluster galaxies at $z<2$ and formation redshifts for such galaxies between $2<z_f<6$ depending on the choice of diagnostics used in its estimation. Furthermore, red sequence galaxies in clusters show a consistent small spread in optical colours up to at least $z\sim2$ and even higher, and hence have  SEDs that were dominated by passively evolving populations since then (\citealt{glazebrook2017}).

Given that the C09 models shown in Fig. \ref{fig:yeps} have $z_f=4$, this would only allow the age of the population to realistically be $\sim1$ Gyr younger or older if the formation redshift was allowed to range between $2<z_f<6$, which would only change the $NUV-r$ colour by $\sim0.1$. This change in $NUV-r$ caused by the age of the population is once again clearly insufficient to account for the total spread in colour observed in our clusters. While the $NUV-r$ colour can be affected by the metallicity and age of the main sequence population, for the cluster galaxy population studied here, variation in this colour must still be dominated by the hot HB stars that are believed to be the primary driving force behind the upturn.

It should be noted that the strength and onset of the upturn also depends on  Reimers mass loss parameter (\citealt{reimers1975,reimers1977}), $\eta$, as the properties of a star on the HB are influenced by the mass of its surviving envelope (i.e. the higher the mass loss during the  RGB phase, the higher the surface temperature of the star in the HB). The YEPS models used here assume $\eta=0.63$, which is calibrated using the inner-halo globular clusters of the Milky Way. However, observations by \cite{dalessandro2012} have shown the existence of a few Milky Way globular clusters such as 47 Tuc which have UV-optical colours much redder than those predicted by the YEPS models with the lowest $Y$, despite having lower metallicities than the models shown here. This discrepancy can be accounted for by assuming a lower value of $\eta$, which would make the UV-optical colours redder. The majority of our galaxies have $NUV-r$ colours that fall within the range predicted by the YEPS models, with the reddest ones being accounted for by higher metallicity models, implying that the $\eta\sim 0.6$ is not unreasonable for most of these galaxies. Nevertheless,  the value chosen for $\eta$ could bring an uncertainty of the order $\sim$1 Gyr to the age of the models of a given metallicity.

While there is an appreciable population of low metallicity Galactic globular clusters that demonstrate upturns (either with or without He-enhanced HB populations), there are few nearby observational models for systems with high $Z$ (commensurate with early type cluster galaxies) and high $Y$. As such, we use the YEPS models in Fig. \ref{fig:yeps} to make plausible estimates of $Y=0.40\sim0.41$ and $z_f=4\sim6$ for the helium-enhanced sub-populations in our observed cluster galaxies. If so, Fig. \ref{fig:yeps} implies that the strength of such upturns should diminish in higher redshift cluster populations because there is insufficient time to form the most He-enriched HBs \citep[see][et seq.]{brown1997}. To a limited extent (given the available time), this  could be countered by an even earlier formation redshift. Clearly, the strength of the evolution in UV upturns at higher redshifts is a crucial diagnostic of the viability of the helium-enhanced HB explanation for these upturns and as such we will explore this issue  in an observational analysis  of higher redshift clusters in a future paper.

While we have focused on explaining our results through the existence of a He-enhanced HB population in these galaxies, the observed lack of evolution over this time interval is still consistent with alternative models for the UV upturn, such as the binary model of \cite{han2007} or a metallicity-dependent mass loss fraction on the red giant branch. It is less consistent with the predictions of  the low-metallicity model proposed by \cite{yi1998}. Again, observations of higher redshift clusters should help  to discriminate between these models.

\section{Conclusions}

By determining the range of UV to optical colours displayed by the population of red sequence galaxies drawn from HST observations of Abell~1689 at $z=0.18$, we demonstrate that it is comparable to that seen for the same population in lower redshift clusters including Coma.

The origin of this range is therefore likely to be the same in these clusters and is due to the variation in the strength of UV upturn seen in their galaxies. In common with previous work, we can interpret this  UV upturn as originating from a metal rich He-enhanced horizontal branch population of variable strength in each galaxy.

Given this interpretation, the lack of variation in the range of colours across $\sim 2.2$~Gyr of lookback time can constrain a combination of the level of Helium enhancement in the stellar subpopulation and the time since formation of the population  and galaxies. In particular the results imply $Y\geqslant 0.40$ for a formation redshift of $z_f\sim 4$ and significantly higher for  later formation. Earlier formation still requires a reasonable level of helium enhancement ($Y\sim 0.38$).

The results cannot currently rule out other origins for the UV upturns, but subsequent observations of red sequence populations in higher redshift clusters should test all of these scenarios.

\section*{Acknowledgements}

SSA is funded by an STFC PhD studentship and thanks the University of Turku and FINCA for their hospitality and local funding during the visits where part of this work was carried out.  This work was based on observations made with the NASA/ESA Hubble Space Telescope, obtained from the data archive at the Space Telescope Science Institute. STScI is operated by the Association of Universities for Research in Astronomy, Inc. under NASA contract NAS 5-26555.

\bibliographystyle{yahapj}
\bibliography{references}

\begin{thebibliography}{}
\providecommand\natexlab[1]{#1}
\providecommand\JournalTitle[1]{#1}

\bibitem[{{Alavi} {et~al.}(2016){Alavi}, {Siana}, {Richard}, {Rafelski},
  {Jauzac}, {Limousin}, {Freeman}, {Scarlata}, {Robertson}, {Stark}, {Teplitz},
  \& {Desai}}]{alavi2016}
{Alavi}, A., {Siana}, B., {Richard}, J., {et~al.} 2016,
  \href{http://dx.doi.org/10.3847/0004-637X/832/1/56}{\JournalTitle{\apj}, 832,
  56}

\bibitem[{{Ali} {et~al.}(2018){Ali}, {Bremer}, {Phillipps}, \& {De
  Propris}}]{ali2018}
{Ali}, S.~S., {Bremer}, M.~N., {Phillipps}, S., \& {De Propris}, R. 2018,
  \href{http://dx.doi.org/10.1093/mnras/sty227}{\JournalTitle{\mnras}, 476,
  1010}

\bibitem[{{Avila}(2017)}]{avila2017}
{Avila}, R.~J. 2017, {Advanced Camera for Surveys Instrument Handbook for Cycle
  25 v. 16.0}

\bibitem[{{Ba{\~n}ados} {et~al.}(2010){Ba{\~n}ados}, {Hung}, {De Propris}, \&
  {West}}]{banados2010}
{Ba{\~n}ados}, E., {Hung}, L.-W., {De Propris}, R., \& {West}, M.~J. 2010,
  \href{http://dx.doi.org/10.1088/2041-8205/721/1/L14}{\JournalTitle{\apjl},
  721, L14}

\bibitem[{{Badenes} {et~al.}(2018){Badenes}, {Mazzola}, {Thompson}, {Covey},
  {Freeman}, {Walker}, {Moe}, {Troup}, {Nidever}, {Allende Prieto}, {Andrews},
  {Barb{\'a}}, {Beers}, {Bovy}, {Carlberg}, {De Lee}, {Johnson}, {Lewis},
  {Majewski}, {Pinsonneault}, {Sobeck}, {Stassun}, {Stringfellow}, \&
  {Zasowski}}]{badenes2017}
{Badenes}, C., {Mazzola}, C., {Thompson}, T.~A., {et~al.} 2018,
  \href{http://dx.doi.org/10.3847/1538-4357/aaa765}{\JournalTitle{\apj}, 854,
  147}

\bibitem[{{Bertin} \& {Arnouts}(1996)}]{bertin1996}
{Bertin}, E., \& {Arnouts}, S. 1996,
  \href{http://dx.doi.org/10.1051/aas:1996164}{\JournalTitle{\aaps}, 117, 393}

\bibitem[{{Bertola} {et~al.}(1982){Bertola}, {Capaccioli}, \&
  {Oke}}]{bertola1982}
{Bertola}, F., {Capaccioli}, M., \& {Oke}, J.~B. 1982,
  \href{http://dx.doi.org/10.1086/159758}{\JournalTitle{\apj}, 254, 494}

\bibitem[{{Boissier} {et~al.}(2018){Boissier}, {Cucciati}, {Boselli}, {Mei}, \&
  {Ferrarese}}]{boissier2018}
{Boissier}, S., {Cucciati}, O., {Boselli}, A., {Mei}, S., \& {Ferrarese}, L.
  2018, \JournalTitle{ArXiv e-prints},
  \href{http://arxiv.org/abs/1801.00985}{{\sffamily arXiv:1801.00985}}

\bibitem[{{Boselli} {et~al.}(2005){Boselli}, {Cortese}, {Deharveng}, {Gavazzi},
  {Yi}, {Gil de Paz}, {Seibert}, {Boissier}, {Donas}, {Lee}, {Madore},
  {Martin}, {Rich}, \& {Sohn}}]{boselli2005}
{Boselli}, A., {Cortese}, L., {Deharveng}, J.~M., {et~al.} 2005,
  \href{http://dx.doi.org/10.1086/444534}{\JournalTitle{\apjl}, 629, L29}

\bibitem[{{Bressan} {et~al.}(1994){Bressan}, {Chiosi}, \&
  {Fagotto}}]{bressan1994}
{Bressan}, A., {Chiosi}, C., \& {Fagotto}, F. 1994,
  \href{http://dx.doi.org/10.1086/192073}{\JournalTitle{\apjs}, 94, 63}

\bibitem[{{Brown} {et~al.}(2000{\natexlab{a}}){Brown}, {Bowers}, {Kimble}, \&
  {Ferguson}}]{brown2000a}
{Brown}, T.~M., {Bowers}, C.~W., {Kimble}, R.~A., \& {Ferguson}, H.~C.
  2000{\natexlab{a}},
  \href{http://dx.doi.org/10.1086/312466}{\JournalTitle{\apjl}, 529, L89}

\bibitem[{{Brown} {et~al.}(2000{\natexlab{b}}){Brown}, {Bowers}, {Kimble},
  {Sweigart}, \& {Ferguson}}]{brown2000b}
{Brown}, T.~M., {Bowers}, C.~W., {Kimble}, R.~A., {Sweigart}, A.~V., \&
  {Ferguson}, H.~C. 2000{\natexlab{b}},
  \href{http://dx.doi.org/10.1086/308566}{\JournalTitle{\apj}, 532, 308}

\bibitem[{{Brown} {et~al.}(1997){Brown}, {Ferguson}, {Davidsen}, \&
  {Dorman}}]{brown1997}
{Brown}, T.~M., {Ferguson}, H.~C., {Davidsen}, A.~F., \& {Dorman}, B. 1997,
  \href{http://dx.doi.org/10.1086/304187}{\JournalTitle{\apj}, 482, 685}

\bibitem[{{Brown} {et~al.}(1998{\natexlab{a}}){Brown}, {Ferguson}, {Deharveng},
  \& {Jedrzejewski}}]{brown1998a}
{Brown}, T.~M., {Ferguson}, H.~C., {Deharveng}, J.-M., \& {Jedrzejewski}, R.~I.
  1998{\natexlab{a}},
  \href{http://dx.doi.org/10.1086/311743}{\JournalTitle{\apjl}, 508, L139}

\bibitem[{{Brown} {et~al.}(2003){Brown}, {Ferguson}, {Smith}, {Bowers},
  {Kimble}, {Renzini}, \& {Rich}}]{brown2003}
{Brown}, T.~M., {Ferguson}, H.~C., {Smith}, E., {et~al.} 2003,
  \href{http://dx.doi.org/10.1086/374035}{\JournalTitle{\apjl}, 584, L69}

\bibitem[{{Brown} {et~al.}(1998{\natexlab{b}}){Brown}, {Ferguson}, {Stanford},
  \& {Deharveng}}]{brown1998b}
{Brown}, T.~M., {Ferguson}, H.~C., {Stanford}, S.~A., \& {Deharveng}, J.-M.
  1998{\natexlab{b}},
  \href{http://dx.doi.org/10.1086/306079}{\JournalTitle{\apj}, 504, 113}

\bibitem[{{Burstein} {et~al.}(1988){Burstein}, {Bertola}, {Buson}, {Faber}, \&
  {Lauer}}]{burstein1988}
{Burstein}, D., {Bertola}, F., {Buson}, L.~M., {Faber}, S.~M., \& {Lauer},
  T.~R. 1988, \href{http://dx.doi.org/10.1086/166304}{\JournalTitle{\apj}, 328,
  440}

\bibitem[{{Chung} {et~al.}(2017){Chung}, {Yoon}, \& {Lee}}]{chung2017}
{Chung}, C., {Yoon}, S.-J., \& {Lee}, Y.-W. 2017,
  \href{http://dx.doi.org/10.3847/1538-4357/aa6f19}{\JournalTitle{\apj}, 842,
  91}

\bibitem[{{Code} \& {Welch}(1979)}]{code1979}
{Code}, A.~D., \& {Welch}, G.~A. 1979,
  \href{http://dx.doi.org/10.1086/156825}{\JournalTitle{\apj}, 228, 95}

\bibitem[{{Conroy} {et~al.}(2009){Conroy}, {Gunn}, \& {White}}]{conroy2009}
{Conroy}, C., {Gunn}, J.~E., \& {White}, M. 2009,
  \href{http://dx.doi.org/10.1088/0004-637X/699/1/486}{\JournalTitle{\apj},
  699, 486}

\bibitem[{{Dalessandro} {et~al.}(2012){Dalessandro}, {Schiavon}, {Rood},
  {Ferraro}, {Sohn}, {Lanzoni}, \& {O'Connell}}]{dalessandro2012}
{Dalessandro}, E., {Schiavon}, R.~P., {Rood}, R.~T., {et~al.} 2012,
  \href{http://dx.doi.org/10.1088/0004-6256/144/5/126}{\JournalTitle{\aj}, 144,
  126}

\bibitem[{{De Propris} {et~al.}(2016){De Propris}, {Bremer}, \&
  {Phillipps}}]{depropris2016}
{De Propris}, R., {Bremer}, M.~N., \& {Phillipps}, S. 2016,
  \href{http://dx.doi.org/10.1093/mnras/stw1521}{\JournalTitle{\mnras}, 461,
  4517}

\bibitem[{{Dorman} {et~al.}(1995){Dorman}, {O'Connell}, \& {Rood}}]{dorman1995}
{Dorman}, B., {O'Connell}, R.~W., \& {Rood}, R.~T. 1995,
  \href{http://dx.doi.org/10.1086/175428}{\JournalTitle{\apj}, 442, 105}

\bibitem[{{Dressel}(2017)}]{dressel2017}
{Dressel}, L. 2017, {Wide Field Camera 3 Instrument Handbook for Cycle 25 v.
  9.0}

\bibitem[{{Glazebrook} {et~al.}(2017){Glazebrook}, {Schreiber}, {Labb{\'e}},
  {Nanayakkara}, {Kacprzak}, {Oesch}, {Papovich}, {Spitler}, {Straatman},
  {Tran}, \& {Yuan}}]{glazebrook2017}
{Glazebrook}, K., {Schreiber}, C., {Labb{\'e}}, I., {et~al.} 2017,
  \href{http://dx.doi.org/10.1038/nature21680}{\JournalTitle{\nat}, 544, 71}

\bibitem[{{Gonzaga}(2012)}]{gonzaga2012}
{Gonzaga}, S. 2012, {The DrizzlePac Handbook}

\bibitem[{{Greggio} \& {Renzini}(1990)}]{greggio1990}
{Greggio}, L., \& {Renzini}, A. 1990,
  \href{http://dx.doi.org/10.1086/169384}{\JournalTitle{\apj}, 364, 35}

\bibitem[{{Han} {et~al.}(2007){Han}, {Podsiadlowski}, \&
  {Lynas-Gray}}]{han2007}
{Han}, Z., {Podsiadlowski}, P., \& {Lynas-Gray}, A.~E. 2007,
  \href{http://dx.doi.org/10.1111/j.1365-2966.2007.12151.x}{\JournalTitle{\mnras},
  380, 1098}

\bibitem[{{J{\o}rgensen} {et~al.}(2017){J{\o}rgensen}, {Chiboucas}, {Berkson},
  {Smith}, {Takamiya}, \& {Villaume}}]{jorgensen2017}
{J{\o}rgensen}, I., {Chiboucas}, K., {Berkson}, E., {et~al.} 2017,
  \href{http://dx.doi.org/10.3847/1538-3881/aa96a3}{\JournalTitle{\aj}, 154,
  251}

\bibitem[{{Kron}(1980)}]{kron1980}
{Kron}, R.~G. 1980,
  \href{http://dx.doi.org/10.1086/190669}{\JournalTitle{\apjs}, 43, 305}

\bibitem[{{Lee}(1994)}]{lee1994}
{Lee}, Y.-W. 1994,
  \href{http://dx.doi.org/10.1086/187451}{\JournalTitle{\apjl}, 430, L113}

\bibitem[{{Mieske} {et~al.}(2004){Mieske}, {Infante}, {Ben{\'{\i}}tez}, {Coe},
  {Blakeslee}, {Zekser}, {Ford}, {Broadhurst}, {Illingworth}, {Hartig},
  {Clampin}, {Ardila}, {Bartko}, {Bouwens}, {Brown}, {Burrows}, {Cheng},
  {Cross}, {Feldman}, {Franx}, {Golimowski}, {Goto}, {Gronwall}, {Holden},
  {Homeier}, {Kimble}, {Krist}, {Lesser}, {Martel}, {Menanteau}, {Meurer},
  {Miley}, {Postman}, {Rosati}, {Sirianni}, {Sparks}, {Tran}, {Tsvetanov},
  {White}, \& {Zheng}}]{mieske2004}
{Mieske}, S., {Infante}, L., {Ben{\'{\i}}tez}, N., {et~al.} 2004,
  \href{http://dx.doi.org/10.1086/423701}{\JournalTitle{\aj}, 128, 1529}

\bibitem[{{Miglio} {et~al.}(2012){Miglio}, {Brogaard}, {Stello}, {Chaplin},
  {D'Antona}, {Montalb{\'a}n}, {Basu}, {Bressan}, {Grundahl}, {Pinsonneault},
  {Serenelli}, {Elsworth}, {Hekker}, {Kallinger}, {Mosser}, {Ventura},
  {Bonanno}, {Noels}, {Silva Aguirre}, {Szabo}, {Li}, {McCauliff}, {Middour},
  \& {Kjeldsen}}]{miglio2012}
{Miglio}, A., {Brogaard}, K., {Stello}, D., {et~al.} 2012,
  \href{http://dx.doi.org/10.1111/j.1365-2966.2011.19859.x}{\JournalTitle{\mnras},
  419, 2077}

\bibitem[{{Newman} {et~al.}(2014){Newman}, {Ellis}, {Andreon}, {Treu},
  {Raichoor}, \& {Trinchieri}}]{newman2014}
{Newman}, A.~B., {Ellis}, R.~S., {Andreon}, S., {et~al.} 2014,
  \href{http://dx.doi.org/10.1088/0004-637X/788/1/51}{\JournalTitle{\apj}, 788,
  51}

\bibitem[{{Park} \& {Lee}(1997)}]{park1997}
{Park}, J.-H., \& {Lee}, Y.-W. 1997,
  \href{http://dx.doi.org/10.1086/303594}{\JournalTitle{\apj}, 476, 28}

\bibitem[{{Piotto} {et~al.}(2005){Piotto}, {Villanova}, {Bedin}, {Gratton},
  {Cassisi}, {Momany}, {Recio-Blanco}, {Lucatello}, {Anderson}, {King},
  {Pietrinferni}, \& {Carraro}}]{piotto2005}
{Piotto}, G., {Villanova}, S., {Bedin}, L.~R., {et~al.} 2005,
  \href{http://dx.doi.org/10.1086/427796}{\JournalTitle{\apj}, 621, 777}

\bibitem[{{Piotto} {et~al.}(2007){Piotto}, {Bedin}, {Anderson}, {King},
  {Cassisi}, {Milone}, {Villanova}, {Pietrinferni}, \& {Renzini}}]{piotto2007}
{Piotto}, G., {Bedin}, L.~R., {Anderson}, J., {et~al.} 2007,
  \href{http://dx.doi.org/10.1086/518503}{\JournalTitle{\apjl}, 661, L53}

\bibitem[{{Price} {et~al.}(2011){Price}, {Phillipps}, {Huxor}, {Smith}, \&
  {Lucey}}]{price2011}
{Price}, J., {Phillipps}, S., {Huxor}, A., {Smith}, R.~J., \& {Lucey}, J.~R.
  2011,
  \href{http://dx.doi.org/10.1111/j.1365-2966.2010.17862.x}{\JournalTitle{\mnras},
  411, 2558}

\bibitem[{{Ree} {et~al.}(2007){Ree}, {Lee}, {Yi}, {Yoon}, {Rich}, {Deharveng},
  {Sohn}, {Kaviraj}, {Rhee}, {Sheen}, {Schawinski}, {Rey}, {Boselli}, {Rhee},
  {Donas}, {Seibert}, {Wyder}, {Barlow}, {Bianchi}, {Forster}, {Friedman},
  {Heckman}, {Madore}, {Martin}, {Milliard}, {Morrissey}, {Neff},
  {Schiminovich}, {Small}, {Szalay}, \& {Welsh}}]{ree2007}
{Ree}, C.~H., {Lee}, Y.-W., {Yi}, S.~K., {et~al.} 2007,
  \href{http://dx.doi.org/10.1086/518125}{\JournalTitle{\apjs}, 173, 607}

\bibitem[{{Reimers}(1975)}]{reimers1975}
{Reimers}, D. 1975, \JournalTitle{Memoires of the Societe Royale des Sciences
  de Liege}, 8, 369

\bibitem[{{Reimers}(1977)}]{reimers1977}
---. 1977, \JournalTitle{\aap}, 61, 217

\bibitem[{{Salaris} {et~al.}(2016){Salaris}, {Cassisi}, \&
  {Pietrinferni}}]{salaris2016}
{Salaris}, M., {Cassisi}, S., \& {Pietrinferni}, A. 2016,
  \href{http://dx.doi.org/10.1051/0004-6361/201628181}{\JournalTitle{\aap},
  590, A64}

\bibitem[{{Schlafly} \& {Finkbeiner}(2011)}]{schlafly2011}
{Schlafly}, E.~F., \& {Finkbeiner}, D.~P. 2011,
  \href{http://dx.doi.org/10.1088/0004-637X/737/2/103}{\JournalTitle{\apj},
  737, 103}

\bibitem[{{Schombert}(2016)}]{schombert2016}
{Schombert}, J.~M. 2016,
  \href{http://dx.doi.org/10.3847/0004-6256/152/6/214}{\JournalTitle{\aj}, 152,
  214}

\bibitem[{{Smith} {et~al.}(2012){Smith}, {Lucey}, \& {Carter}}]{smith2012}
{Smith}, R.~J., {Lucey}, J.~R., \& {Carter}, D. 2012,
  \href{http://dx.doi.org/10.1111/j.1365-2966.2012.20524.x}{\JournalTitle{\mnras},
  421, 2982}

\bibitem[{{Tailo} {et~al.}(2017){Tailo}, {D'Antona}, {Milone}, {Bellini},
  {Ventura}, {Di Criscienzo}, {Cassisi}, {Piotto}, {Salaris}, {Brown},
  {Vesperini}, {Bedin}, {Marino}, {Nardiello}, \& {Anderson}}]{tailo2017}
{Tailo}, M., {D'Antona}, F., {Milone}, A.~P., {et~al.} 2017,
  \href{http://dx.doi.org/10.1093/mnras/stw2790}{\JournalTitle{\mnras}, 465,
  1046}

\bibitem[{{Tantalo} {et~al.}(1996){Tantalo}, {Chiosi}, {Bressan}, \&
  {Fagotto}}]{tantalo1996}
{Tantalo}, R., {Chiosi}, C., {Bressan}, A., \& {Fagotto}, F. 1996,
  \JournalTitle{\aap}, 311, 361

\bibitem[{{Yi} {et~al.}(1997){Yi}, {Demarque}, \& {Oemler}}]{yi1997}
{Yi}, S., {Demarque}, P., \& {Oemler}, Jr., A. 1997,
  \href{http://dx.doi.org/10.1086/304498}{\JournalTitle{\apj}, 486, 201}

\bibitem[{{Yi} {et~al.}(1998){Yi}, {Demarque}, \& {Oemler}}]{yi1998}
---. 1998, \href{http://dx.doi.org/10.1086/305078}{\JournalTitle{\apj}, 492,
  480}

\bibitem[{{Yi} {et~al.}(1999){Yi}, {Lee}, {Woo}, {Park}, {Demarque}, \&
  {Oemler}}]{yi1999}
{Yi}, S., {Lee}, Y.-W., {Woo}, J.-H., {et~al.} 1999,
  \href{http://dx.doi.org/10.1086/306856}{\JournalTitle{\apj}, 513, 128}

\end{thebibliography}

\label{lastpage}
\end{document}